\documentclass[doublespace, 12pt fonts]{aastex6}
\usepackage{amsmath}
\usepackage{yhmath}

\shorttitle{Is FRB 200428 an FRB?}
\shortauthors{Du \& Wang}

\begin{document}

\title{What can we learn from FRB 200428?}

\author{Shuang Du$^{1,2}$, Weihua Wang$^{1,2}$}
\affil{
$^{1}${State Key Laboratory of Nuclear Physics and Technology, School of Physics, Peking University, Beijing 100871, China}\\
$^{2}${Kavli Institute for Astronomy and Astrophysics, Peking University, Beijing 100871, China}}

\email{dushuang@pku.edu.cn}

\begin{abstract}
The two radio pulses (TRPs) from SGR 1935+2154 detected by \cite{BKRMHB} and \cite{2020arXiv200510324T}
have similar features to that of cosmological fast radio bursts (FRBs).
Many authors directly call the TRPs as FRB 200428 without consider two questions carefully.
(1) Are the TRPs just two brighter subpulses of a normal radio pulse liking the normal radio pulses seen in other magnetars during their outburst?
(2) If the TRPs are two bursts of an FRB, does this make other FRBs difficult to understand?
In this paper, we try to clarify these two questions.
First, we compare these TRPs with previous observations of normal radio pulses from pulsars and magnetars,
and find that the TRPs should be produced by different mechanism from that of normal radio pulses.
We then investigate the second question by assuming the TRPs have the same origin as that of periodically repeating FRBs
and find that the origin of the periodicity of periodically repeating FRBs should not be induced by precession.
Otherwise, we should expect that the TRPs are not an FRB, at least the TRPs don't have the same origin as that of periodically repeating FRBs.
\end{abstract}


\section{Introduction}\label{sec1}
Fast radio bursts (FRBs) are millisecond-duration bright radio transients \citep{2007Sci...318..777L,2013Sci...341...53T,2016Natur.531..202S}.
Their origins remain highly debated, but neutron stars (NSs) are widely considered as their sources (see \citealt{2018PrPNP.103....1K,2019A&ARv..27....4P} for reviews).
The strong association between FRB 121102 and persistent radio and optical counterparts \citep{2017Natur.541...58C,MPH}
suggests FRB sources could be young NSs born in supernovae \citep{MBM}.
The detection of periodically repeating FRBs indicates that  FRBs could arise from precessing NSs or NSs in binaries \citep{2020Natur.582..351C,2020arXiv200303596R}.

\cite{BKRMHB} and \cite{2020arXiv200510324T} report the detection of two bright radio pulses (hereafter, TRPs) from SGR 1935+2154 during the outbursts of this magnetar \citep{2020arXiv200511071L,2020ApJ...898L..29M,2020arXiv200511178R,2020arXiv200512164T}.
These TRPs are directly named FRB 200428 (see, e.g., \citealt{2020ApJ...897L..40D,2020MNRAS.499.2319K})
due to their short intrinsic durations $\sim 0.60 \;\rm ms$ and $\sim 0.34 \;\rm ms$ (with separation $\sim 29\;\rm ms$)
and high average fluence $\sim 700\;\rm kJy\; ms$ and $\sim 1.5\;\rm MJy\; ms$ \citep{BKRMHB,2020arXiv200510324T}.
Since SGR 1935+2154 is a magnetar with rotational period $3.24\;\rm s$, spin-down rate $1.43\times 10^{-11}\;\rm s\cdot s^{-1}$, inferred effective surface dipole magnetic field strength
$2.2 \times 10^{14}\;\rm G$ and spin-down age $3.6 \;\rm kyr$ \citep{Israel}, and locates in the Galactic supernova remnant
G57.2+0.8 \citep{Gaen}, the sources and origins (i.e., bursts of magnetars) of repeating FRBs\footnote{Nonrepeating FRBs may not have the same origin as that of repeating FRBs} seem to be determined.

Note that, unlike cosmological FRBs, these TRPs are associated with X-ray bursts
and have much lower isotropic-equivalent energy releases. Although the absence of X-ray burst-associated and low-luminosity cosmological FRBs can be naturally attributed to their
greater distance, one can't rule out one other possibility that the TRPs have different origin from that of cosmological FRBs.
For example, the TRPs may be two brighter radio pulses just liking the normal radio pulses seen in other magnetars during their outbursts.

In this paper, we will try to clarify that whether these TRPs are two bursts of an FRB.
In Section 2, we compare the features of the TRPs with normal radio pulses from radio pulsars and magnetars.
In Section 3, we search for the periodic modulation due to the spin of an NS of FRB 121102 again,
since if FRB 121102 has the same origin as that of the TRPs there should be a spin-induced periodic modulation.
In Section 4, we study the origin of nonspin-origin periodicity of periodically repeating FRBs.
In Section 5, we discuss the geometry problem for the precessing NS scenario of periodically repeating FRBs.
In Section 6, we investigate the effect of burst reaction to the SGR-burst associated FRBs.
Section 7 is summary and discussion.
Throughout this paper, we will use the NS parameters that the radius $R_{\ast}=10^{6}\;\rm cm$,
rotational inertia $I=10^{45}\;\rm g\cdot cm^{2}$ and mass $M_{\rm ns}=1.4\;\rm M_{\odot}$ for calculation.

\section{Compare the TRPs with normal radio pulses}

Phenomenologically, these TRPs are different from previous radio pulses from pulsars and magnetars. We list some aspects as follows.
For SGR 1935+2154:
(i) Never have we seen such energetic radio bursts from the other NSs;
(ii) There is frequency up-drift;
(iii) Time interval between these two radio bursts is $\sim 29\;\rm ms$.
For the rest NSs:
(a) Magnetars show long-lasting continuous radio emission during the outbursts but not isolated two bright radio pulses \citep{2014ApJS..212....6O};
(b) No frequency drifting has been detected, even if the pulsar/magnetar shows glitches or outbursts;
(c) The rotational period of SGR 1935+2154 is $3.24\;\rm s$, so the time interval of $\sim 29\;\rm ms$ between the TRPs is not the rotational period.
Therefore, we conclude from these differences that the TRPs
are emitted from a position (or produced by a mechanism) different from that of normal periodic radio pulses of pulsars/magnetars,
i.e., the TRPs are not normal radio pulses and even not two subpulses of a normal radio pulse.

However, being not the normal radio pulses does not necessarily means that the TRPs are two bursts of an FRB.
If the TRPs are just two bursts of an FRB and produced by a certain mechanism of the magnetar just liking that of cosmological FRBs,
one can expect that the repeating FRBs (e.g., FRB 121102) should be modulated by NS spins,
so that a periodicity between successive pulses of repeating FRBs should be found.

\section{Search for rotational period of an NS in FRB 121102}\label{sec3}
\begin{figure}
\centering
  \includegraphics[width=0.45\textwidth]{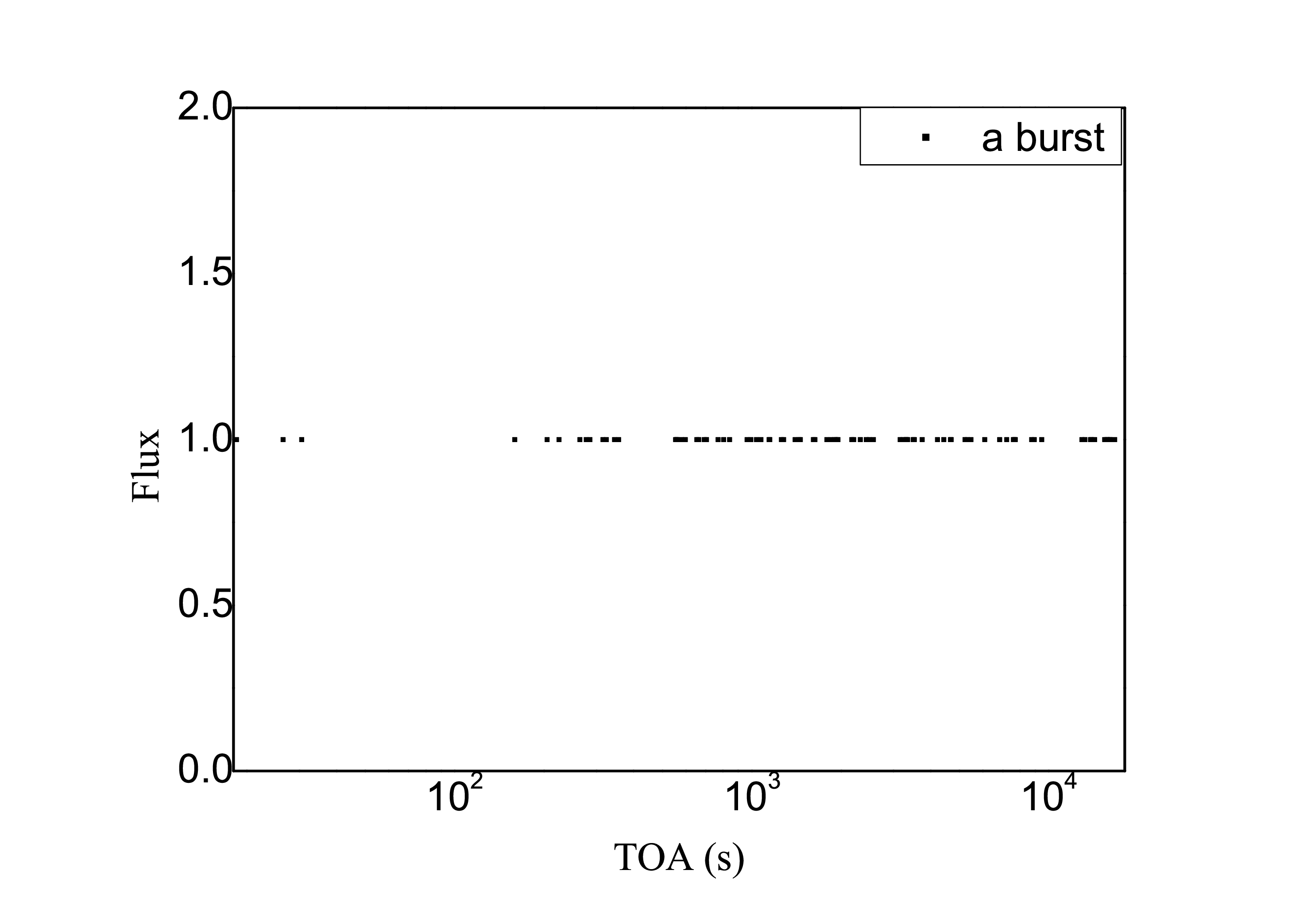}
  \includegraphics[width=0.45\textwidth]{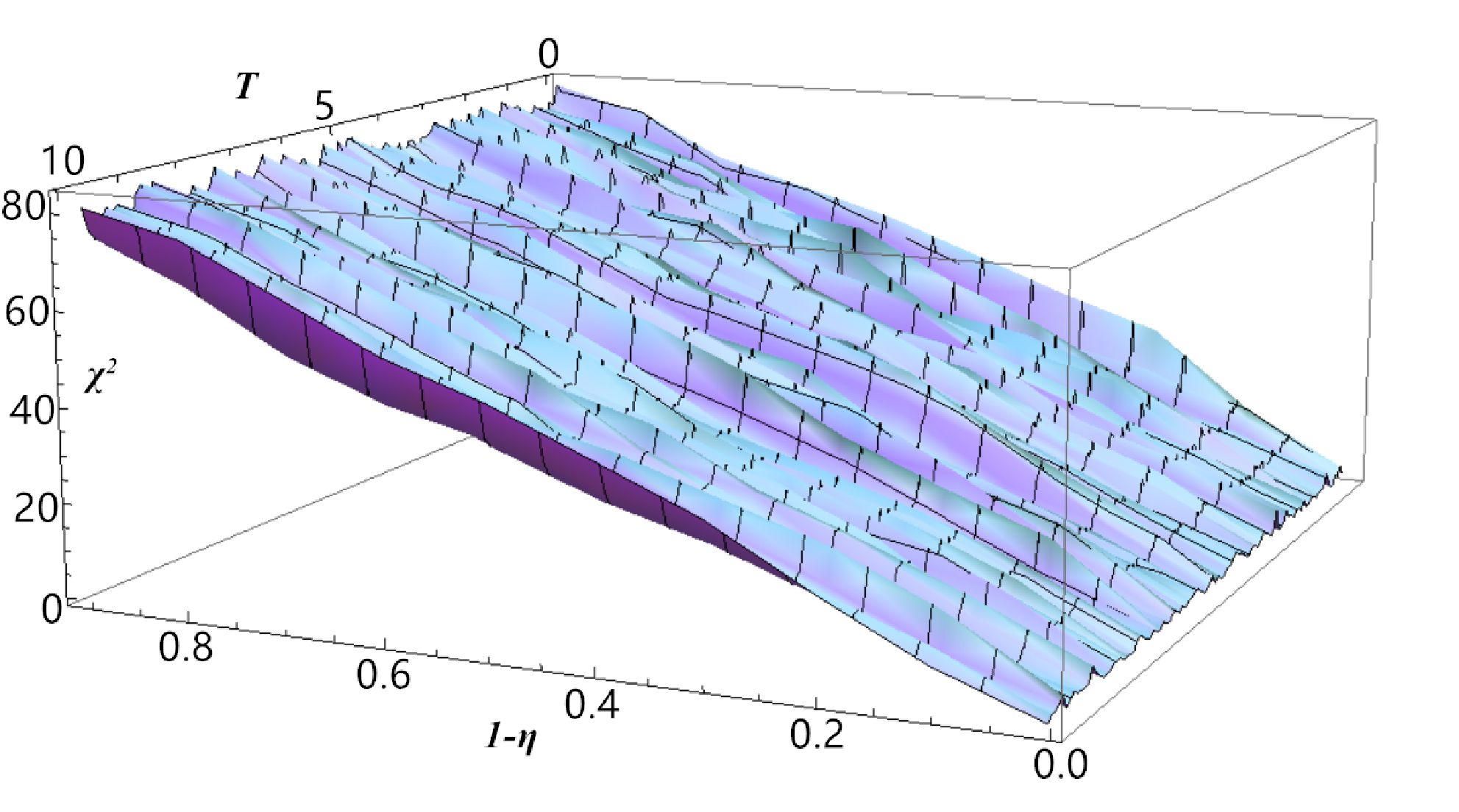}
  \caption{Method. Left panel: Every point means a burst. If there is a period between successive bursts of FRB 121102,
  all the points should locate on the ``high level" of a certain square wave function. Right panel:
  As we can see, $\chi^{2}$ only can achieve $0$ when $\eta$ achieves $1$.
  It indicates that there is no reliable rotational period of an NS in FRB 121102 since $\eta\rightarrow 1$ means that
  the total duration of ``low level" of the square wave function is negligible during $\Delta t$
  and the ``equivalent" period of the function is longer than $\Delta t$.\label{fig1}}
\end{figure}

\cite{2018ApJ...863....2G} and \cite{2018ApJ...866..149Z} reported 93 bursts of FRB 121102 during $\Delta t\sim 16600\;\rm s$ (see Table 2 in \citealt{2018ApJ...866..149Z}).
\cite{2018ApJ...866..149Z} performed a Fourier-domain acceleration and a time-domain acceleration search and no period of an NS was found.
We use their data to search for the rotational period of the NS with another visualized method.
The idea is that the rotational period of the NS only can determine the time when we can see the burst but not the flux of the burst.
Therefore, we only care about the burst even itself but not its flux since it has nothing to do with the rotation of the NS.
Corresponding to every pulse time of arrival $t_{\rm i}$ (with $i=1, 2, ..., 93$ ),
we set the flux of the burst $F(t_{\rm i})$ as a constant, e.g., $1$, so this constant just means that the NS is in an active window.
Then we can use a square wave function $Y(t)$ (with $t$ being the time since the observation) to fit the points in upper panel of Figure \ref{fig1}.
If FRB 121102 is really modulated by the NS rotation, there should be a square wave function with certain duty cycle $\eta$ and period $T$
can make all the points locate on the ``high level", i.e.,
\begin{eqnarray}\label{A1}
\chi^{2}=\sum_{i=1}^{93} [F(t_{\rm i})-Y(t_{\rm i})]^{2}=0.
\end{eqnarray}

Since we have assumed FRB 121102 originates from a magnetar just like the TRPs,
before the fitting, we should set the range of the rotational period of the magnetar.
Given that a newborn magnetar may be fast rotating, the spin-down would be effective at the early stage.
With ignoring the evolution of the magnetic field, the spin-down is
\begin{eqnarray}\label{A2}
\frac{4\pi^{2} I}{P^{2}}\dot{P}=\frac{8\pi^{4}B_{\rm eff}^{2}R_{\ast}^{6} }{3c^{3}P^{4}},
\end{eqnarray}
where $P$ and $B_{\rm eff}$ are the rotational period and effective surface dipole magnetic field strength (include all the deviations from the dipole field) of the magnetar, respectively, and $c$ is the speed of light.
Through equation (\ref{A2}), one has
\begin{eqnarray}\label{A3}
P(t)\simeq 0.04\times\left ( \frac{B_{\rm eff}}{10^{14}\;\rm G} \right )^{2/3}\left ( \frac{t}{1\;\rm yr} \right )^{1/3}\;\rm s.
\end{eqnarray}
Therefore, we search for the the rotational period of the magnetar from $0.1\;\rm s$ to $10\;\rm s$ based on our computing power.

As shown in the lower panel of Figure \ref{fig1}, $\chi^{2}\rightarrow 0$ only can achieve when $\eta\rightarrow 1$.
This is a trivial result since $\eta\rightarrow 1$ means the ``equivalent" period of the square wave function is longer than the duration of the observation ($\sim 16600\;\rm s$),
so that all data points locate on ``high level''.
Therefore no credible periodic modulation due to the NS spin exists in FRB 121102.
This indicates FRB 121102 should not generated by the internal factor of the NS (e.g., SGR bursts) unless the emission beam of FRB 121102 always points the earth.

\section{Where the periods of periodically repeating FRBs come from?}\label{sec4}
According to the above section, if FRB 121102 is induced by the magnetar bursts,
the absence of the modulation due to the NS spin should be attributed to that the emission beam of FRB 121102 always points the earth during its active phase.
Although this is a minor probability situation, we still further consider this situation, i.e., discuss the compatibility between the magnetar-burst-associated TRPs and the
periodically repeating FRBs.
In this paper, we will focus on the precessing scenario of periodically repeating FRBs
since the scenario that the orbital-motion-induced periodicity of periodically repeating FRBs has been discussed in \citep{DWWX}.

The precession of an NS may originate from three cases: (1) spin-orbit coupling induced precession \citep{DR,BO},
(2) forced precession by a fallback disk \citep{QC,QXXW}, (3) free/radiative precession (see, e.g., \citealt{1969mech.book.....L}).
The spin-orbit coupling only can induce a spin-precession period which is much longer than that of periodically repeating FRBs,
even for the known most compact relativistic system PSR J0737-3039 (order of $\sim100\;\rm yr$; \citealt{MB,AGL}).
Besides, no hint shows that SGR 1935+2154 belongs to a binary system. Actually, all the detected magetars are isolated.
So case (1) should be ruled out.

Under the fallback-disk-forced precession model \citep{TWW},  the precession period is
\begin{eqnarray}\label{T1}
P_{\rm pre}\sim 100\left ( \frac{M_{\theta}}{10^{-5}\;\rm M_{\odot}} \right )^{-1}\left ( \frac{P}{5\;\rm ms} \right )^{3}\;\rm day,
\end{eqnarray}
where $M_{\theta}$ is the mass of accretion disk which makes contribution to the precession.
Under this model, the magnetar must be young, so that there is enough fallback matter.
Assuming the kick velocity of the magnetar is low, so that the fallback disk can keep stable.
The fallback matter (e.g., supernova ejecta) usually has the codirectional angular momentum as that of its progenitor, as well as the remnant NS,
i.e., the direction of the angular momentum of the fallback disk is usually parallel to the spin vector of the NS, so that $M_{\theta}$ should be much smaller that the total disk mass.
This is different from the accretion between an NS and the companion stellar, since the angular momentums of the two stars are usually misalignment.
Therefore, experimentally, a fallback disk with considerable precession is hard to appear.
On the other hand, before the formation of the fallback disk,
the fallback ejecta should approximately fall back in a spherically symmetric way as the ejecta falls back to the vicinity of the NS from a far away place.
To form a disk near the NS under this model, the ram pressure of fallback matter should be larger than that of the spin-down wind.
The outward pressure of spin-down wind is directly proportional to $r^{-2}$ with $r$ being the distance from the acting point to the center of the NS,
and the inward ram pressure of fallback matter is directly proportional to $r^{-5/2}$  \citep{PC}.
Therefore, the ram pressure of fallback matter should strong enough at lest at the corotation radius ($r_{\rm co}=(GM_{\rm ns}/\Omega^{2})^{1/3}$), i.e.,
\begin{eqnarray}\label{eq7}
\frac{\dot{m}}{8\pi}\left ( \frac{2GM_{\rm ns}}{r_{\rm co}^{5}} \right )^{1/2}\geq \frac{2\pi^{3}B_{\rm eff}^{2}R_{\ast}^{6} }{3c^{4}P^{4}r_{\rm co}^{2}},
\end{eqnarray}
where $\dot{m}$ is the fallback accretion rate.
From equation (\ref{eq7}), there is
\begin{eqnarray}
\dot{m}\geq2.0\times 10^{-2}\left ( \frac{B_{\rm eff}}{10^{14}\;\rm G} \right )^{2}\left ( \frac{P}{5\;\rm ms} \right )^{-11/3}\;\rm M_{\odot }\cdot yr^{-1}.
\end{eqnarray}
This accretion rate demands the total mass of the disk to be at least $\sim 0.1 \;\rm M_{\odot }$. If one demands the disk matter can be accreted onto the NS surface,
the total mass of the disk will be much larger and the corresponding magnetar must be very young (see also \citealt{MBM}).
However, observation indicates that the source of periodically repeating FRB 180916.J0158+65 is $800\;\rm kyr$ to $7 \;\rm Myr$ old \citep{TPK}.
Besides, being in a high-accretion-rate state for a long time many lead to the decay of the magnetic field strength of the NS,
so that a magnetar-like FRB source (e.g., SGR 1935+2154) is hard to appear.
However, according to equation (\ref{T1}), if one increases $P$ by $4$ times, $M_{\theta}$ should be increased by $160$ times and the total mass of the disk must be even larger,
such a massive disk only can live for a very short time (at most $\sim 1000\;\rm s$ as indicated by the durations of gamma-ray bursts).
In a word, a fallback disk which satisfies equation (\ref{T1}) should not be long lived and frequent\footnote{
It is worth mentioning that the fallback accretion discussed here is different from that of accretion-powered X-ray pulsars.
One also can note that being in a high-accretion-rate state (the mass of the disk $\gg10^{-5}\;\rm M_{\odot}$) would effect the evolution of $P$ effectively,
as well as the precession period, so this model also can be tested by examining the evolution of the FRB periodicity.}.

Under the free/radiative precession models \citep{2020arXiv200204595L,2020ApJ...892L..15Z,DNS},
the relationship between free precession period and rotational period is $P_{\rm pre}\propto P/\varepsilon$ with $\varepsilon$  being the ellipticity of the NS.
So the older the NS is, the longer the precession it has.
However observations do not support this speculation that the NS corresponding to FRB 180916 should be older \citep{2017Natur.541...58C,MPH,MNH,2018Natur.553..182M}
under the isolated-NS scenario \citep{MBM}.

\section{Difficulties encountered by all precession scenarios: the geometry problem.}
\begin{figure}
\centering
  \includegraphics[width=0.4\textwidth]{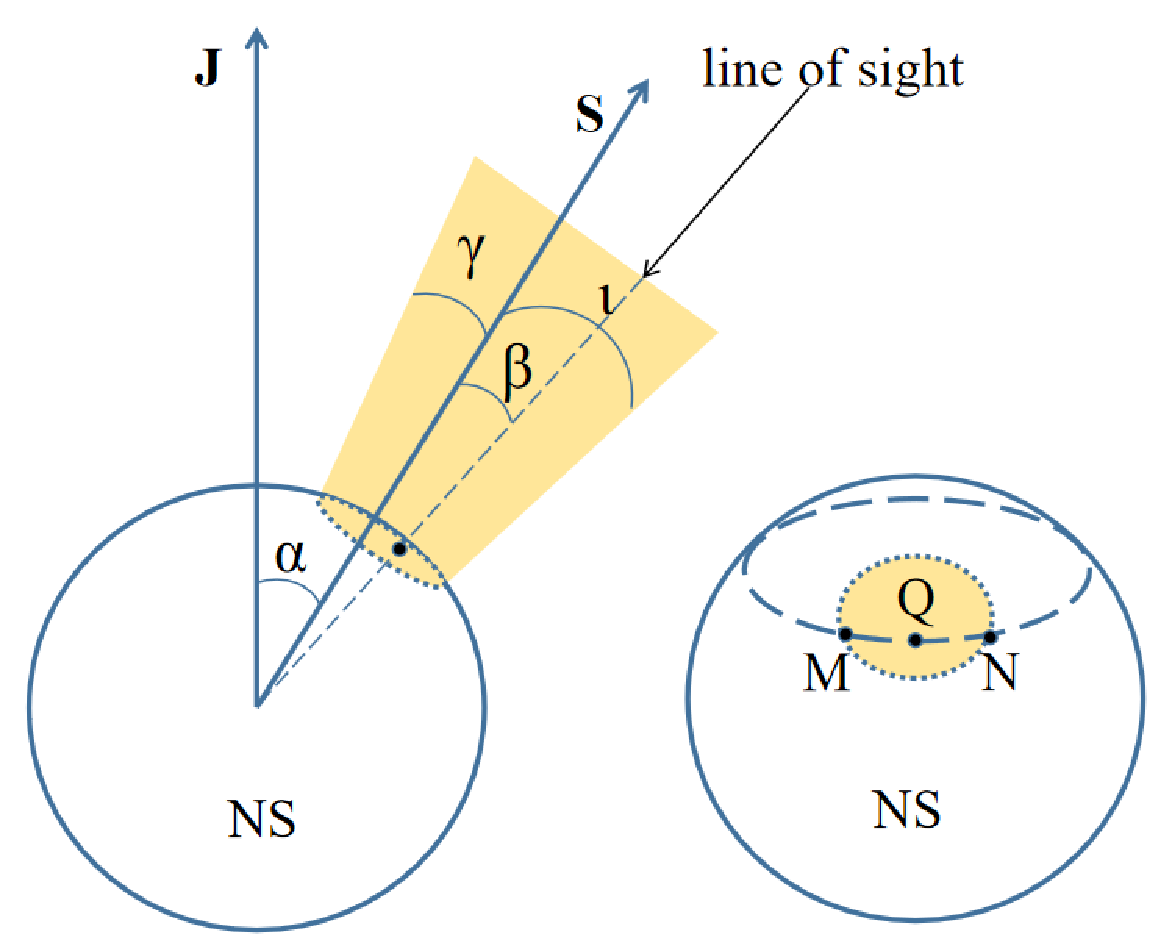}
  \caption{The geometry of the precession scenario, where $\mathbf{S}$ is the spin vector and $\mathbf{J}$ is the total angular momentum.
  Since the FRB beam always points the earth during the NS spin, there is $\beta <\gamma$.
  The precession will induce the opening and closing of the FRB, so that there is $\iota<\alpha$.
  The total relation is $\beta <\gamma <\iota <\alpha$. The intersection of the line of sight and the NS surface forms a circumference $\wideparen{NQMN}$ (dashed ellipse) due to the precession.
  The part of $\wideparen{NQMN}$ in the projection (dotted green ellipse) of FRB beam onto the NS surface is arc $\wideparen{NQM}$.\label{fig3}}
\end{figure}
Aside from the problems discussed above, all the precession scenarios have the same difficult:
the rotation around the spin angular momentum $\mathbf{S}$ can not lead the FRB beam to be a gyroscope-like radio beam,
but the precession of the spin vector around the total angular momentum $\mathbf{J}$ would result in the switching on/off of the FRB beam.
This demands a strange geometry.
For a rotating NS, its total angular momentum should be mainly contributed by the spin angular momentum,
so the wobble angle $\alpha\sim \left | \mathbf{J}-\mathbf{S} \right |/\left | \mathbf{J} \right |$ should be very small .
As shown in Figure \ref{fig3}, there is $\iota <\alpha$, so that the opening angle of the FRB beam ($<2\iota$) should also be small.
However, to reproduce the observed duty cycles for FRB 180916.J0158+65 ($\eta\approx 0.25$; \citealt{2020Natur.582..351C}) and FRB 121102 ($\eta\approx 0.5$; \citealt{2020arXiv200303596R}),
the size scales of FRB beams should satisfy
\begin{eqnarray}
\frac{\wideparen{NQM}}{\wideparen{NQMN}}=\eta.
\end{eqnarray}
Therefore, the size scale of the radiation region must be very large. This contradicts the previous discussion that the opening angles of FRB beams should be very small.
Besides, no matter the FRB emission comes from the open field line region \citep{WZCX} or the close field line region \citep{2016ApJ...829...27D},
curved field lines always make the opening angle of the FRB emission should not be too small.
We must expect that $\alpha$ is large enough (e.g., larger than the opening angler of open field line region
if the FRB is produced in this region), but this an unusual request since we have never seen such distinct precession in about 2600 pulsars.
In addition, the precessing FRB models demand a fixed angle (position) between the FRB beam and total NS angular momentum $\mathbf{J}$,
this further demands the associated accidental SGR bursts to always occur at a fixed position (it is too optimistic).

\section{the effect of the burst reaction: the change of the period of precession}
Suggesting that the burst of an FRB is associated with an SGR burst just like the TRPs.
Note that the reaction of an SGR burst on the NS should not be strictly centripetal, an extra torque will exert on the NS (see Appendix for more details).
The upper limit of the change in angular momentum can be estimated as
\begin{eqnarray}\label{eq1}
\Delta J_{\rm max}=\frac{E_{\rm burst}}{c}r=3.3\times 10^{40} \left ( \frac{E_{\rm burst}}{10^{45}\;\rm erg} \right )\;\rm erg\cdot s,
\end{eqnarray}
where $E_{\rm burst}$ is the total energy release of the burst (see, e.g., SGR 1806-20, \citealt{2005Natur.434.1098H}).
Correspondingly, the upper limit on change of rotation frequency is
\begin{eqnarray}\label{eq2}
|\Delta\omega_{\rm max}|=\frac{\Delta J_{\rm max}}{2\pi I}=5.3\times 10^{-6} \left ( \frac{E_{\rm burst}}{10^{45}\;\rm erg} \right )\;\rm rad\; s^{-1};
\end{eqnarray}
and the upper limit on induced angular velocity of precession is
\begin{eqnarray}\label{eq3}
\dot{\varphi}_{\rm max}&\sim& \frac{rPL_{\rm burst}}{2\pi Ic}\nonumber\\
&=&5.3\times 10^{-6}\left ( \frac{L_{\rm burst}}{10^{45}\;\rm erg\; s^{-1}} \right )
\left(\frac{P}{1\;\rm s}\right)\;\rm rad\; s^{-1},
\end{eqnarray}
where $L_{\rm burst}$ is the luminosity of the burst.
Optimistically, from equation (\ref{eq3}), during an energetic SGR burst, a forced precession with a period $\sim 10-100\;\rm day$ could be excited.
Besides, a free precession will be induced after the SGR burst (see equation (\ref{eq24})).
Therefore, the period of an FRB induced by free/radiative precession would be changed due to an energetic SGR burst ($\omega_{z}'$ would be changed, see equation (\ref{eq16'}) ).
If the TRPs are two bursts of an FRB and associated with the ``interior-factor''-induced SGR busts,
the period of repeating FRBs could not be interpreted reasonably as long as some of the SGR bursts are energetic.
However, radiation reaction could be one of the sources of a glitch (see equation (\ref{eq2})).

Although burst reaction can not be the origin of periods of repeating FRBs,
equation (\ref{eq3}) shows a modulation on $t_{\rm i}$, e.g.,
\begin{eqnarray}\label{eqa1}
t_{93}-t_{93}'<\frac{\Delta t}{P}\cdot \frac{2\pi }{\omega^{2}}|\Delta\omega_{\rm max}|=2.6\times 10^{-3} \left ( \frac{P}{1\;\rm s} \right ) \;\rm s,
\end{eqnarray}
where $t_{93}'$ is the theoretical pulse time of arrival of the 93rd burst without the modulation by radiation reaction.
From equation (\ref{eqa1}), it is clear to see that $t_{93}-t_{93}'\ll \eta T$ can be easily satisfied,
so that the accumulated modulation would not effect the search of rotational period since $\Delta t$ is short enough.
The result shown in Section 3 is reasonable.

\section{Summary and discussion}

According to the above discussion, the realistic celestial bodies can not provide the physical conditions that needed by these precession scenarios.
If we treat the TRPs to be a repeating FRB and assume the TRPs have the same origin as that of periodically repeating FRBs,
the origin of the periodicity of periodically repeating FRBs should not be induced by precession.
As discussed in \citep{DWWX}, the periodicity should be induced by an external factor.
Otherwise, we should expect that the TRPs are not bursts of an FRB, at least the TRPs don't have the same origin as that of periodically repeating FRBs.
It seems that the pulsar-asteroid/asteroid belt impact model \citep{2015ApJ...809...24G,2016ApJ...829...27D} can reconcile the contradiction
among nonrepeating FRBs (NSs collide with stray asteroids), nonperiodically repeating FRBs (NSs with parabolic orbits) and
periodically repeating FRBs (NSs collide with asteroid belts) well.
Under this model, the phase delay detected by \cite{P1} and \cite{P2} can be explained as:
the massive asteroids collide with the NS at first, massive asteroids can penetrate the NS magnetosphere to produce a high-frequency repeating FRB;
the smaller asteroids captured by the NS will collide with the NS whereafter, small asteroids can not get into the interior of the magnetosphere, so that the delayed bursts will be low-frequency.

\section{Acknowledgement}
We acknowledge useful discussion at the pulsar group of PKU.
We thank Drs. Heng Xu and Weiyang Wang for useful comments which greatly enhance the stringency of this paper.
This work was supported by the National Key R\&D Program of China (Grant
No. 2017YFA0402602), the National Natural Science Foundation of China (Grant Nos. 11673002, and U1531243), and
the Strategic Priority Research Program of Chinese Academy Sciences (Grant No. XDB23010200).

\section{Appendix}\label{app1}
\begin{figure}
\centering
  \includegraphics[width=0.4\textwidth]{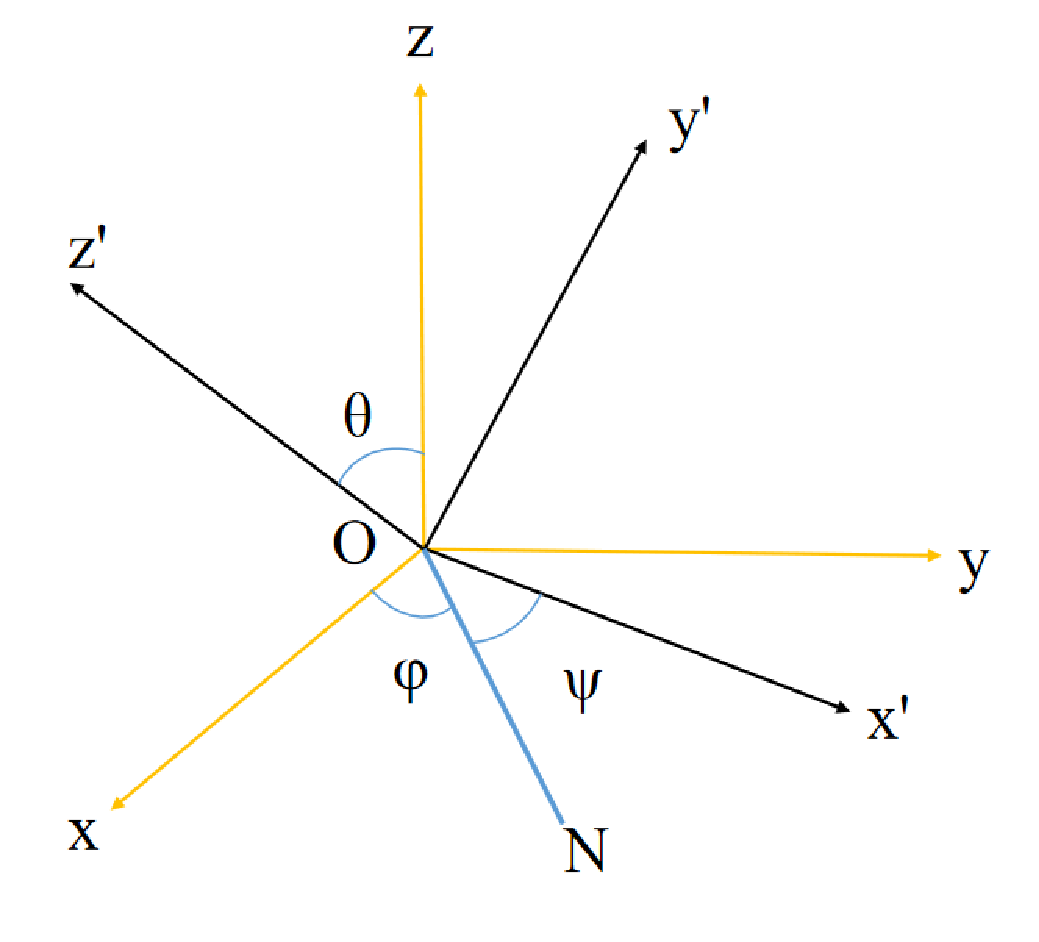}
  \caption{The geometry between the frame of principal axis of inertia $Ox'y'z'$ and lab frame $Oxyz$.
  $\varphi$ (precession angle), $\psi$  (rotation angle), $\theta$ (nutation angle) are three Euler angles.
  $\overline{ON}$ is the intersecting line between the plane $Ox'y'$ and plane $Oxy$. \label{fig2}}
\end{figure}

Let's consider this problem more analytically.
We treat an NS as an Euler gyro that the rotational inertias along the $x'$-axis and $y'$-axis are the same (see Figure \ref{fig2}), i.e., $I_{11}'=I_{22}'$.
The rotation of this NS is determined by Euler's dynamics equations (in the frame $Ox'y'z'$; see e.g., \citealt{1969mech.book.....L})
\begin{eqnarray}\label{eq4}
I_{11}^{\prime}\dot{\omega }_{\rm x}^{\prime}-(I_{22}^{\prime}-I_{33}^{\prime})\omega _{\rm y}^{\prime}\omega _{z}^{\prime}=M_{\rm x}^{\prime},
\end{eqnarray}
\begin{eqnarray}\label{eq5}
I_{22}^{\prime}\dot{\omega}_{\rm y}^{\prime}-(I_{33}^{\prime}-I_{11}^{\prime})\omega _{z}^{\prime}\omega _{\rm x}^{\prime}=M_{\rm y}^{\prime},
\end{eqnarray}
\begin{eqnarray}\label{eq6}
I_{33}^{\prime}\dot{\omega}_{\rm z}^{\prime}=M_{\rm z}^{\prime},
\end{eqnarray}
where
\begin{eqnarray}
M_{\rm i}'=\eta_{\rm i}\frac{L_{\rm burst}}{c}r, (i=x,y,z)
\end{eqnarray}
is the component of the torque induced by radiation reaction with $\eta_{\rm i}\in (0,1)$ being the contribution of the burst luminosity to the torque
(note that $\eta_{\rm x}$, $\eta_{\rm y}$ and $\eta_{\rm z}$ compete with each other),
and $\omega_{\rm i}'$ is the component of the spin vector.

Under the initial conditions that $\omega_{\rm x}'(t=0)=\omega_{\rm y}'(t=0)=0$, $\omega_{0}=\omega_{\rm z}'(t=0)$, and the assumption that
\begin{eqnarray}
L_{\rm burst}=\begin{cases}
L_{\rm p}\left ( \frac{t}{t_{\rm p}} \right )^{\alpha } & \text{ if } 0<t<t_{\rm p}, \\
L_{\rm p}\left ( \frac{t}{t_{\rm p}} \right )^{\beta } & \text{ if } t_{\rm p}<t<t_{\rm e},
\end{cases}
\end{eqnarray}
with $L_{\rm p}$ being the peak luminosity, $t_{\rm p}$ being the peak time and $t_{\rm e}$ being the duration time of the burst, we may solve the above equations (\ref{eq4})-(\ref{eq6}).
From equation (\ref{eq6}), there is
\begin{eqnarray}
\omega_{\rm z}'(t)=\begin{cases}
\frac{\eta_{\rm z}L_{\rm p}r}{I_{33}' c}\frac{t^{\alpha +1}}{(\alpha +1)t_{\rm p}^{\alpha }}+\omega_{0} & \text{ if } 0<t<t_{\rm p}, \\
\frac{\eta_{\rm z}L_{\rm p}r}{I_{33}' c}\left [\frac{t^{\beta  +1}}{(\beta  +1)t_{\rm p}^{\beta}}+ \frac{t_{\rm p}}{\alpha +1} \right ]+\omega_{0}& \text{ if } t_{\rm p}<t<t_{\rm e}.
\end{cases}
\end{eqnarray}
From equations (\ref{eq4}) and (\ref{eq5}), we have
\begin{eqnarray}\label{eq13}
\ddot{\omega }_{\rm x}'-\frac{\dot{\omega}_{\rm z}'}{\omega_{\rm z}'}\dot{\omega}_{\rm x}'+\varepsilon^{2}\omega_{\rm z}'^{2}\omega_{\rm x}' =\frac{\dot{M}_{\rm x}'}{I_{11}'}-\varepsilon\frac{ M_{\rm y}'}{I_{22}'}\omega_{\rm z}'-\frac{M_{\rm x}'\dot{\omega }_{\rm z}'}{I_{11}'\omega_{z}'},
\end{eqnarray}
and
\begin{eqnarray}\label{eq14}
\ddot{\omega }_{\rm y}'-\frac{\dot{\omega}_{\rm z}'}{\omega_{\rm z}'}\dot{\omega}_{\rm y}'+\varepsilon^{2}\omega_{\rm z}'^{2}\omega_{\rm y}' =\frac{\dot{M}_{\rm y}'}{I_{22}'}+\varepsilon\frac{ M_{\rm y}'}{I_{11}'}\omega_{\rm z}'-\frac{M_{\rm y}'\dot{\omega }_{\rm z}'}{I_{22}'\omega_{z}'},
\end{eqnarray}
where $\varepsilon=(I_{33}'-I_{11}')/I_{22}'$. Therefore, equations (\ref{eq13}) and (\ref{eq14}) can be solved numerically for the given $\eta_{\rm i}$ and $\varepsilon$.
Since the value of $\varepsilon$ should be very small, we simplify the discussion as
\begin{eqnarray}\label{eq15}
I_{11}^{\prime}\dot{\omega }_{\rm x}^{\prime}\approx M_{\rm x}^{\prime},
\end{eqnarray}
\begin{eqnarray}\label{eq16}
I_{22}^{\prime}\dot{\omega}_{\rm y}^{\prime}\approx M_{\rm y}^{\prime},
\end{eqnarray}
so that there are
\begin{eqnarray}
\omega_{\rm x}'(t)=\begin{cases}
\frac{\eta_{\rm x}L_{\rm p}r}{I_{11}' c}\frac{t^{\alpha +1}}{(\alpha +1)t_{\rm p}^{\alpha }}& \text{ if } 0<t<t_{\rm p}, \\
\frac{\eta_{\rm x}L_{\rm p}r}{I_{11}' c}\left [\frac{t^{\beta  +1}}{(\beta  +1)t_{\rm p}^{\beta}}+ \frac{t_{\rm p}}{\alpha +1} \right ]& \text{ if } t_{\rm p}<t<t_{\rm e},
\end{cases}
\end{eqnarray}
and
\begin{eqnarray}
\omega_{\rm y}'(t)=\begin{cases}
\frac{\eta_{\rm y}L_{\rm p}r}{I_{22}' c}\frac{t^{\alpha +1}}{(\alpha +1)t_{\rm p}^{\alpha }}& \text{ if } 0<t<t_{\rm p}, \\
\frac{\eta_{\rm y}L_{\rm p}r}{I_{22}' c}\left [\frac{t^{\beta  +1}}{(\beta  +1)t_{\rm p}^{\beta}}+ \frac{t_{\rm p}}{\alpha +1} \right ]& \text{ if } t_{\rm p}<t<t_{\rm e}.
\end{cases}
\end{eqnarray}
Clearly, $\omega=\sqrt{\Sigma_{i=x,y,z}\omega_{\rm i}'^{2}}$ is not a constant, one can imagine that precession and nutation should exist simultaneously.

After the burst, we input the zeroth approximation
\begin{eqnarray}\label{eq16'}
\begin{cases}
\omega_{\rm x}'(t=t_{\rm e})=\frac{\eta_{\rm x}L_{\rm p}r}{I_{11}' c}\left [\frac{t_{\rm e}^{\beta +1}}{(\beta  +1)t_{\rm p}^{\beta}}+ \frac{t_{\rm p}}{\alpha +1} \right ]\\
\omega_{\rm y}'(t=t_{\rm e})=\frac{\eta_{\rm y}L_{\rm p}r}{I_{22}' c}\left [\frac{t_{\rm e}^{\beta +1}}{(\beta  +1)t_{\rm p}^{\beta}}+ \frac{t_{\rm p}}{\alpha +1} \right ] \\
\omega_{\rm z}'(t=t_{\rm e})=\frac{\eta_{\rm z}L_{\rm p}r}{I_{33}' c}\left [\frac{t_{\rm e}^{\beta +1}}{(\beta  +1)t_{\rm p}^{\beta}}+ \frac{t_{\rm p}}{\alpha +1} \right ]+\omega_{0}
\end{cases}
\end{eqnarray}
into the homogeneous Euler's dynamics equations as the new initial conditions.
The solution under $t>t_{\rm e}$ is
\begin{eqnarray}\label{eq17}
\begin{cases}
\omega_{\rm x}'(t)=A\cos[\varepsilon\omega_{\rm z}'(t=t_{\rm e})t+\phi_{0}]\\
\omega_{\rm y}'(t)=A\sin[\varepsilon\omega_{\rm z}'(t=t_{\rm e})t+\phi_{0}] \\
\omega_{\rm z}'(t)=\omega_{\rm z}'(t=t_{\rm e})
\end{cases}
\end{eqnarray}
where
\begin{eqnarray}
\phi_{0}=\arctan\left ( \frac{\omega_{\rm y}'(t=t_{\rm e})}{\omega_{\rm x}'(t=t_{\rm e})} \right )-\varepsilon \omega_{\rm z}'(t=t_{\rm e})t_{\rm e},
\end{eqnarray}
\begin{eqnarray}
A=\frac{\omega_{\rm x}'(t=t_{\rm e})}{\cos[\varepsilon \omega_{\rm z}'(t=t_{\rm e})t_{\rm e}+\phi_{0}]}.
\end{eqnarray}
After the burst, $\omega=\sqrt{\Sigma_{i=x,y,z}\omega_{\rm i}'^{2}}$ is a constant, only the precession should be left.
The evolution of the three Euler angles can be easily solved through the conservation of angular momentum (e.g.,\citealt{1969mech.book.....L}),
i.e.,
\begin{eqnarray}\label{eq20}
J\sin\theta \sin\psi =I_{11}^{\prime}A\cos(\varepsilon\omega_{\rm z}'(t=t_{\rm e}) t+\phi _{0}),
\end{eqnarray}
\begin{eqnarray}
J\sin\theta \cos\psi =I_{11}^{\prime}A\sin(\varepsilon\omega_{\rm z}'(t=t_{\rm e}) t+\phi _{0}),
\end{eqnarray}
\begin{eqnarray}\label{eq22}
J\cos\theta =I_{33}^{\prime}\omega _{z}^{\prime},
\end{eqnarray}
where
\begin{eqnarray}
J=\sqrt{2I_{11}'^{2}A^{2}+I_{33}'^{2}\omega_{\rm z}'^{2}(t=t_{\rm e})}
\end{eqnarray}
is the total angular momentum.

Combining equations (\ref{eq20})-(\ref{eq22}) gives
\begin{eqnarray}\label{eq24}
\dot\varphi=\omega_{\rm z}'(t=t_{\rm e})(1+\varepsilon)\sec\theta_{0},
\end{eqnarray}
where
\begin{eqnarray}\label{eq25}
\theta_{0}=\arccos\left(\frac{I_{33}'\omega_{\rm z}'(t=t_{\rm e})}{J}\right).
\end{eqnarray}
So the precession period can be estimated as $P_{\rm pre}={2\pi}/{\dot\varphi}$.

\end{document}